\documentclass{iopart}
\usepackage{epsfig}

\def\jou#1#2#3#4#5{#5 {#1} $\!\!$#2 {\bf #3} #4}

\newfam\BMath
\font\BMathL=cmmib10 
\font\BMathl=cmmib7
\font\BMathm=cmmib5
\textfont\BMath=\BMathL \scriptfont\BMath=\BMathl
\scriptscriptfont\BMath=\BMathm

\renewcommand\P{{\fam\BMath p}}

\def\O{\Omega}
\def\aO{\bar \Omega}

\def\a{\alpha}
\def\c{\chi}

\def\f{\phi}

\def\l{\lambda} 
\def\m{\mu}
\def\n{\nu}

\def\p{\pi}

\def\r{\rho}
\def\s{\sigma}
\def\t{\tau}

\def\cl{\mathcal L}

\def\cp{\mathcal P}

\def\del{\partial}
\def\tr{\mbox{\rm tr}}
\def\exp{\mbox{\rm exp}}

\def\ra{\rightarrow}

\def\bc{\begin{center}}
\def\ec{\end{center}}
\def\be{\begin{equation}}
\def\ee{\end{equation}}
\def\bea{\begin{eqnarray*}}
\def\eea{\end{eqnarray*}}
\def\bfi{\begin{figure}}
\def\efi{\end{figure}} 

\begin{document}

\title[Anomalous Production of $\O$ and $\aO$ is Evidence $\dots$] 
{Anomalous Production of $\O$ and $\aO$ is Evidence 
 for the Formation of Disoriented Chiral Condensates?}

\author{J I Kapusta\dag\ and S M H Wong\ddag  
\footnote[3]{Speaker}}

\address{\dag\ School of Physics and Astronomy, University of Minnesota,
Minneapolis, \\ MN 55455, USA}

\address{\ddag\ Physics Department, Ohio State University, 
Columbus, OH 43210, USA}

\begin{abstract}
No conventional picture of nucleus-nucleus collisions has yet been 
able to explain the abundance of $\O$ and $\aO$ hadrons in 
central collisions between Pb nuclei at 158 A GeV at the CERN SPS.  
We argue that this is evidence that they are produced from
topological defects in the form of skyrmions arising from the 
formation of disoriented chiral condensates. It is shown
that the excess $\O$ or $\aO$ produced could not be easily
washed out in the hadronic medium and therefore would survive 
in the final state. 
\end{abstract} 




\section{Introduction}
\label{s:intro}

$\O$ and $\aO$ are very peculiar baryons in that they are
very hard to manufacture from hadron based material. 
Doubly strange baryons will first have to be made from two singly
strange hadrons via hadronic scattering before they can collide
with another singly strange hadron to make the triply strange
$\O$. As such $\O$ and $\aO$ require a very long time 
to come into chemical equilibrium in a hadron gas. It is a virtual
impossibility in nuclear collisions if only a hadron gas 
is formed in such collisions \cite{kmr}. 

While the yields of other hadrons in heavy ion collisions do not 
present too much of a problem for dynamical models to reproduce, 
this is not true for strange hyperons, especially the $\O$ and $\aO$. 
For example, one of the more well-known and sophisticated models, namely 
the Ultra-relativistic Quantum Molecular Dynamics Model, which is 
basically a hadronic model supplemented with strings for particle production, 
could only generate much lower yields than the experimental measurements 
at the SPS. Only by readjusting a few paramenters in the model, either 
by increasing the string tension by a factor of three or by reducing 
the constituent quark masses down to the values of the current quark masses, 
could the same yield as found at the SPS be reproduced \cite{urqmd}. 
Other dynamical models also have problems in generating the same
triply strange hyperons abundance as found at the SPS. 
The fact that other hadrons did not present much problem for the
various models seemed to suggest that for the majority
of the particles the essence of the basic production mechanisms 
as well as their interactions had been successfully captured in these 
models. Alas, this does not seem to be true for $\O$ and $\aO$.

Further signs of an anomaly can be found in the thermal fit to the
particle ratios as measured in experiments. Such fits are useful in helping
to determine whether the system as a whole has achieved thermal as well 
as chemical equilibrium. A global freeze-out temperature, collective
flow velocity, system size at freeze-out, chemical potentials etc
can be obtained. The quality of the fit is arguably captured in the
value of the $\c^2$. The smaller it is, the better is the fit. 
It was found that if particle ratios as measured by the WA97 and
the NA49 collaborations at the SPS involving $\O$ and $\aO$ were
excluded, then the value of $\c^2$ would be reduced by an order of 
magnitude \cite{rl1,rl2}. The triply strange hyperon and antihyperon 
therefore deviate from the other hadrons. Any suspicion to the 
contrary is dispelled by an examination of the trend of the slopes
of the $m_T$ spectra of the hadrons. Most of the hadrons exhibit 
a linear increase of the value of the slope parameter with increasing
mass \cite{rol,cal,mar}. The $\O$ and $\aO$ show little desire of 
following this linear trend with a deviation of six or more standard 
deviations \cite{san,wa97}. 

In brief $\O$ and $\aO$ show every sign of being a very different
type of hadron. There is an overall tendency of more 
$\O$ and $\aO$ being produced than expected in most cases. This 
seems to suggest that their production is not well understood or 
there might be an additional mechanism that has not been taken into
account. We will propose a mechanism below for this and attempt
to provide arguments, support, and experimental evidence that such a 
possibility is not excluded by the data at the SPS.

\section{Disoriented Chiral Condensates and the Skyrme Model}
\label{s:dcc}

In \Sref{s:intro} we explained why there seemed to exist an anomaly 
in the yield of the $\O$ and $\aO$, and a novel mechanism for
their production was called for. We now make the claim that this
anomaly is caused by the formation of Disoriented Chiral Condensates 
(DCC) \cite{kw}. This claim obviously requires a lot of explanation;  
afterall, DCC are usually associated with low energy pions or more precisely 
the observable for DCC is the distribution of the ratio of neutral to 
overall pion yield and not baryons. In addition DCC have proven to be 
elusive; all searches for them so far ended invariably in failure. 
Therefore the chance that our invocation of the possibility of DCC 
formation is correct might apparently seem rather remote. 
However one should not jump to the conclusion that just because DCC
did not manifest themselves in the distribution of the low energy 
pion ratio, it would necessarily mean that no DCC were formed. It was 
realized quite early that DCC could only be observed if sufficiently 
large domains were formed in heavy ion collisions \cite{rw}. 
In the absence of large domains, the distribution of the low energy 
neutral to total pion ratio could not be distinguished from the case of 
no DCC formation. Provided that our claim does not require that
large domains be formed then there will be no contradiction. 
This will be discussed later when our idea is applied to experimental
data at the SPS. 

Having partially answered the question about the fact that no DCC have 
been observed and stressed that it could not be fully answered in this 
section without considering experimental data, let us turn to the other 
question about how hyperons and antihyperons can be related to DCC. 
Four decades ago Skyrme considered the following Lagrangian density
\be \cl_S = \frac{f_\p^2}{4} \; \tr(\del_\m U\del^\m U^\dagger) 
           +\frac{1}{32g^2}  \; \tr[U^\dagger \del_\m U,U^\dagger \del_\n U]^2
\ee 
where 
\be U = \exp \{i \mbox{\boldmath $\t$} \cdot \f /f_\p\} 
      = ( \s + i \mbox{\boldmath $\t$} \cdot \mbox{\boldmath $\p$})/f_\p 
        \; ,
\ee
$f_\p$ is the pion decay constant and $g$ is a coupling constant
which turned out to be essentially that of the $\p$-$\r$-$\p$ 
interaction. This Lagrangian density is later known as that of the 
Skyrme model. It consists of the non-linear sigma model and the
Skyrme quartic interaction term. The latter is quartic in the sense 
that it has four $U\del_\m U$ units. The equation of motion 
from this Lagrangian is known to have classical topologically 
non-trivial solutions known as skyrmions. They come in the totally 
spherically symmetric form 
\be U = U_S = \exp \{i \mbox{\boldmath $\t$}\cdot \mbox{\boldmath $\hat r$}\; 
                     F(r) \}
\ee
where $F(r)$ must satisfy certain very specific boundary conditions. 
They are 
\be F(r \ra \infty) \ra 0  \mbox{\hspace{.8cm} and \hspace{.8cm}} 
    F(r=0) = N \p \; .
\ee 
$N$ in the last expression is the integral valued winding number. 
It has been identified as the baryon number \cite{sk,w1,w2}. 
These classical solutions are therefore baryons or antibaryons
for $N=\pm 1$. It is clear now that classical chiral fields can
not only produced pions, they can also generate baryons and 
antibaryons. Unlike making pions, however, the classical chiral fields 
must have a non-trivial topology before they can produce baryons. 
Equivalently they must acquire a non-zero winding number in order to
generate skyrmions or anti-skyrmions. This does not happen automatically.

\section{How to get Non-zero Winding Number?}
\label{s:how}

For the classical fields to obtain a non-zero winding number is
not trivial. To describe how this might happen, it is best to 
illustrate it with the linear sigma model whose Lagrangian density is 
\be \cl = \frac{1}{2} \del_\m \Phi^\a \del^\m \Phi_\a 
         -\frac{\l}{4} (\Phi^\a \Phi_\a -v^2 )^2 
         -V_{\mbox{\scriptsize} SB} 
\label{eq:lsm} 
\ee
where $\Phi_\a = (\s,\p_1,\p_2,\p_3)$ is a compactified four component 
field which spans a 4-D space of the chiral fields and 
$V_{\mbox{\scriptsize} SB}$ is the symmetry breaking potential
usually introduced to give the pion a mass and to break the $O(4)$ 
symmetry as manifested by the other terms in \Eref{eq:lsm} to favour the 
$\s$-direction. Our goal here is to convey in simple terms how non-zero
winding number could be obtained from suitable field configurations. 
Including $V_{\mbox{\scriptsize} SB}$ in the discussion will only
complicate matters so in the following we will drop it or set 
$V_{\mbox{\scriptsize} SB}=0$.  

The vacuum of the theory will take centre stage in the following
discussion which, from the second term of \Eref{eq:lsm}, is $S^3$ or the 
surface of a sphere in 4-D. This is also the order parameter space for 
the chiral phase transition or the space of the condensates. 
Since it is hard to visualize a sphere in 4-D, we will temporarily
go back to 3-D in the space of the condensates and to 2-D in actual
space. Once the basic idea is understood, it is easy to extrapolate
back to 4-D. 

Whenever DCC are being considered, one must also include domains in
the discussion. After reducing the dimensions by one everywhere as
mentioned above, the resulting flat space can be triangulated into
domains. An example is shown in \Fref{f:2d} with vertices $P_i$, 
$i=0,\dots, 3$. The size of the domain is such that each vertex 
is separated from the others by at least a unit of the correlation 
length $\xi$ so that each one has, in general, a different value
of the condensate $\a_i$. The latter will be situated somewhere on
$S^2$ the dimensionally reduced order parameter space. This is 
illustrated on the right in \Fref{f:2d}. Forming a winding number of 
one can be crudely understood as having the patch formed by
$\a_1$, $\a_2$ and $\a_3$ in \Fref{f:2d} to cover the whole of 
$S^2$ by stretching them to join $\a_0$ on the other side of the 
sphere. Actually it is not necessary to form full winding all at
once, it is only required for the condensates $\a_i$ to situate and
spread themselves on $S^2$ in such a way that there is a chance for 
full winding to develop in time. The required configuration is that
$\a_0$ must be situated on the antipodal point of the centre of 
the patch formed by $\a_1$, $\a_2$ and $\a_3$ \cite{ks}. By randomly
distributing four points on $S^2$, on the average there is a probability
of $\cp = 1/8$ that such a configuration is formed. The corresponding
triangulated spatial region has an area of $A=3\sqrt{3}\xi^2/4$. 
Therefore the probability per unit area for this to happen is
\be \cp/A = \frac{1}{6\sqrt{3}\xi^2} \;. 
\ee 
\bfi 
  \hspace{3.5cm}
  \epsfig{file=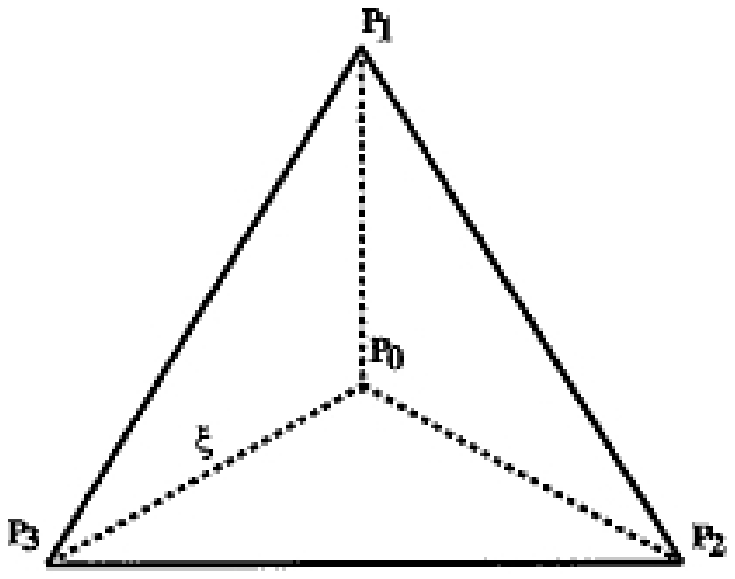,width=4.2cm} \hspace{1cm} 
  \epsfig{file=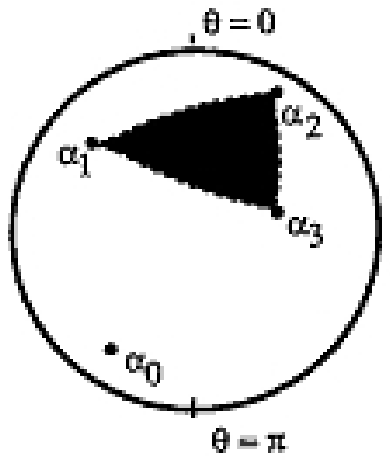,width=3.2cm}  
\caption{A sample domain in 2-D space with vertices $P_i$ which have
         chiral condensates $\a_i$, $i=0,1,2,3$ respectively located 
         on the order parameter space $S^2$.} 
\label{f:2d} 
\efi

Returning to the actual case of the order parameter space being 
$S^3$ and space being 3-dimensional, the triangulated spatial domain 
is now a tetrahedron as shown in \Fref{f:3d} and the ``patch''
formed by the condensates $\a_i$, $i=1,\dots, 4$ on $S^3$ is likewise 
a tetrahedron. The configuration required for full winding to be able
to develop is now that $\a_0$ must be at the ``antipodal'' point of 
the centre of this tetrahedron on $S^3$. Randomly distributing five
points on $S^3$ results in on the average a probability of 
$\cp=1/16$ for this configuration to happen in a spatial volume of
$V=8 \xi^3/9\sqrt{3}$. The probability per unit volume for skyrmion
or anti-skyrmion to form is thus
\be \cp/V \simeq 0.12 \; \xi^{-3}  \;. 
\ee
We have of course discussed an idealized situation where there is no
symmetry breaking, no interactions amongst the neighbouring domains etc. 
When these are taken into account, numerical simulations of global texture 
formation in the expanding universe showed that the actual probability
is smaller by about a factor of three \cite{sp,lee}. Therefore the
total probability per unit volume for the formation of skyrmion
and anti-skyrmion pair (due to conservation of baryon number) is  
\be \cp/V \simeq 0.08 \; \xi^{-3}  \;. 
\ee
Note that the probability per unit volume is larger the smaller the 
domain size. This mechanism is thus orthogonal to the usual observable
of DCC which is the distribution of pion ratio which requires large 
domains to form for observation to be possible. Although the above
probability per unit volume is derived for flavour $SU(2)$, we will 
follow \cite{ek,ek2,eh} and make the identification $SU(2)\simeq SU(3)$
since computations suggested they be similar. 
We are now ready to apply these results to the SPS data. 
\bfi
 \hspace{5cm}
 \epsfig{file=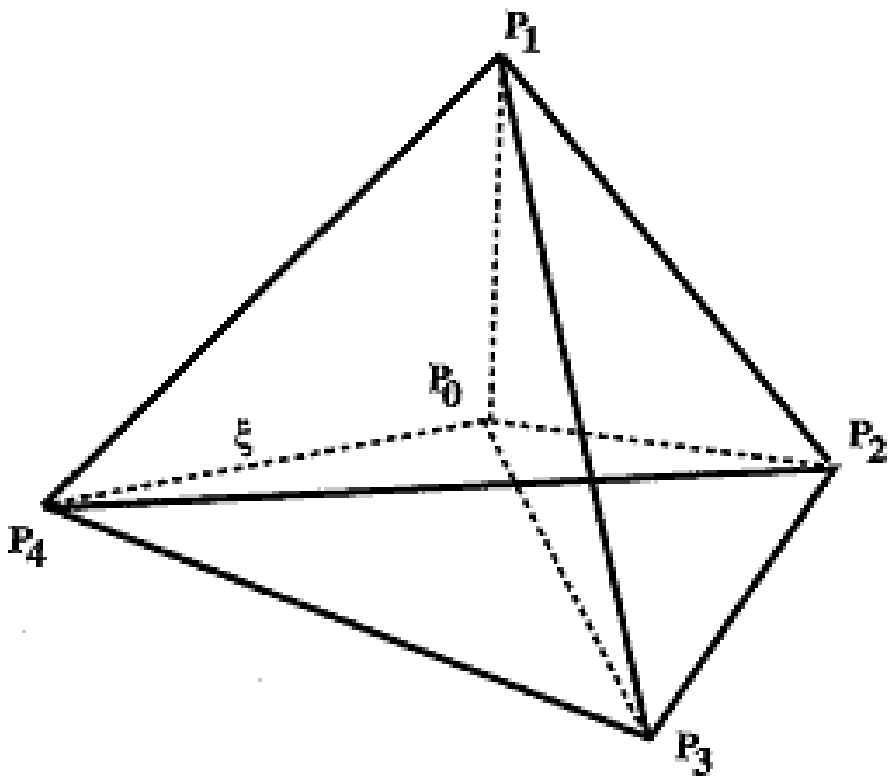,width=5cm} 
\caption{A sample domain in the actual 3-D space whose vertices again 
         have chiral condensates located on different points on $S^3$.} 
\label{f:3d} 
\efi

\section{Skyrmion Formation at the SPS}
\label{s:sps}

Our primary interest is in the $\O$ and $\aO$ yield. At the SPS
data are only available from the WA97 collaboration \cite{cal,san,wa97}. 
Although the NA49 collaboration also measured strange baryon and
antibaryon yields \cite{mar,app,gab}, the results of the triply strange 
baryons are not yet available. However the data of $\O$ and $\aO$
from the WA97 collaboration is only concentrated within one unit 
around $y=0$ because of their limited rapidity coverage. They gave 
\cite{cal} 
\[  \aO/\O  = 0.383 \pm 0.081 \;\;\;\;\;\;\; \mbox{and} \;\;\;\;\;\;\; 
    \O+\aO  = 0.410 \pm 0.08       \;.
\]
The NA49 collaboration has a wider coverage in $y$ but no $\O$ and $\aO$ 
data are available. Fortunately it just so happened that both 
collaborations have measured the yield of the doubly strange $\Xi$ and 
$\bar \Xi$. Furthermore the latter from NA49 have been extrapolated to the 
full momentum space in \cite{bc} for use in thermal analysis. 
From WA97 we have 
\[      \Xi^-  = 1.50 \pm 0.10 \;\;\;\;\;\;\; \mbox{and} \;\;\;\;\;\;\; 
   \bar \Xi^+  = 0.37 \pm 0.06     \;,           
\] 
and from NA49 the extrapolated to full momentum space data are \cite{bc}
\[              \Xi^-  = 7.5  \pm 1.0 \;\;\;\;\;\;\; \mbox{and} \;\;\;\;\;\;\; 
   \Xi^- + \bar \Xi^+  = 8.2  \pm 1.1          \;.  
\]
Collectively such data permit us to extrapolate in turn the
$\O$ and $\aO$ yield to full momentum space. For example 
the total $\aO$ yield can be obtained from the combination 
\be \left (\frac{\aO}{\O+\aO}\right )_{\mbox{\scriptsize WA97}} 
    \left (\frac{\O+\aO}{\Xi^- +\bar \Xi^+} \right )_{\mbox{\scriptsize 
                                                            WA97}} 
    \Big (\Xi^- +\bar \Xi^+ \Big )_{\mbox{\scriptsize NA49}} = 0.498 \;.
\ee
So roughly there can be expected on the average half a $\aO$ per 
central collision at the SPS.

\subsection{The total number of baryon and antibaryon from DCC}
\label{ss:tn}

With the available data, one can also try to estimate the total 
number of baryons and antibaryons originated from DCC. Since
DCC only yield low energy hadrons, it is reasonable to assume that
only octet and decuplet baryons or antibaryons and no higher
resonances will be generated. We also assume that they are equally 
likely to be produced independent of flavour. The fact that there 
are so few $\aO$ prompts us to assume that all of them are from DCC. 
Altogether there are eight spin-1/2 octet baryons and
ten spin-3/2 decuplet baryons which make a total of 16+40=56 
possibilities. Only four of these can yield $\aO$. In order to
give half a $\aO$ per central collision on the average there
must be about seven anti-skyrmions formed per central collision. 
Because of baryon number conservation, anti-skyrmion must form
at the same time as skyrmion so there must be a total of about
fourteen baryons and antibaryons originating from DCC per
central collision. For the more abundant baryons, for example
protons: there are only a quarter of a proton on the average per 
central collision that came from DCC and so this extra source of 
baryons and antibaryons can only be detected from the yields of 
the much rarer hyperons.

\subsection{Domain size}
\label{ss:ds}

In \Sref{s:dcc} we discussed the fact that no DCC have so
far been detected from the distribution of the low energy neutral to 
total pion ratio and that this fact should not automatically be 
interpreted as no DCC formation in heavy ion collisions. 
Rather it could be a sign that no large DCC domains were able to
form because the conditions found in these collisions might not be
favourable for such formations. We now try to reinforce this idea
by estimating the domain size from the available data and from the
theoretical consideration, in particular the probability per unit 
volume for skyrmion and anti-skyrmion formation already discussed 
in the previous sections. 

At the SPS there are about 2000 hadrons being produced in a central
collision. Assuming that DCC formed at a time when the density is
approximately ten times that of nuclear matter density, 
1.7 hadrons/fm$^3$ then the probability $\cp/V \sim 0.08\xi^{-3}$
introduced in \Sref{s:how} would give a domain size of $\xi \sim 2$ fm. 
Theoretical studies in DCC formation gave an estimate of the domain size of 
$\xi \sim 1.5$ fm for a system evolved in time while in equilibrium 
\cite{rw} and a size of $\xi \sim 3-4$ fm in the annealing scenario 
where the effective potential gradually evolved from the chirally
symmetric form back towards the vacuum form \cite{gm}. Our phenomenological
estimate of $\xi \sim 2$ fm is thus consistent with the estimates made 
based on various theoretical considerations. Unfortunately it was known
that domains of such small size would not reveal DCC in the distribution
of the pion ratios. Our point here is that the proposal of DCC as the
extra source of baryons and antibaryons does not contradict the fact
that they have not been observed so far via other means.

\section{Can $\O$ and $\aO$ from DCC Survive?}
\label{s:sur}

So far everything seems to be consistent. However, we must verify that the 
number of $\O$ and $\aO$ thus generated cannot be easily changed 
in the system. If they can be easily destroyed and subsequently 
be recreated then the trace of DCC will be washed out by such processes 
of chemical equilibration. Since chemical equilibration of $\O$ in
a hadronic environment is known to be very slow and inefficient, one
can almost be certain that the triply strange baryons and antibaryons
thus formed are safe. Let us not assume this but make some estimates
as to the timescale required for their destruction or removal from
the system. 

\begin{table}
\caption{The timescales for destruction of $\O$ at different temperatures.} 
\lineup
\begin{tabular}{@{}ccc}
\br
\hspace{.8cm} $T$ [MeV] \hspace{.8cm}  &   
\hspace{.8cm} $\t_{_{\p + \O \rightarrow  K + \Xi}}$ [fm/c] \hspace{.8cm} & 
\hspace{.8cm} $\t_{_{ K + \O \rightarrow \p + \Xi}}$ [fm/c] \hspace{.8cm}   \\ 
\mr
  150        &     500     &    488                            \\
  170        &     370     &    290                            \\
  200        &     257     &    152                            \\ 
\br
\label{t}
\end{tabular}
\end{table} 

Some possible destructive processes are $\O$ being hit by the more
abundant $\p$'s and $K$'s in the system and thus converted into $\Xi$. 
\be \p + \O \longrightarrow  K + \Xi \;\;\;\;\;\;\; \mbox{and} \;\;\;\;\;\;\;  
     K + \O \longrightarrow \p + \Xi    
\ee 
In order to make estimates of the timescale for such process, we will
assume that everything else except the triply strange baryons are in 
equilibrium and the interaction matrix elements between the hadrons are 
universal. Then a rate equation for the destruction of $\O$ or $\aO$ can 
be obtained similarly to the rate equations for chemical equilibration 
as found in \cite{kmr}. We write
\be \frac{\partial \r_\O}{dt} = 
   -\langle \s_{_{  K + \Xi \rightarrow \p + \O}} v_{\p\O} \rangle R_5 
            \r_\p^\infty \r_{_\O} 
   -\langle \s_{_{ \p + \Xi \rightarrow  K + \O}} v_{ K\O} \rangle R_1 
            \r_K^\infty  \r_{_\O} 
\label{eq:rd}
\ee
where 
\be R_5 = {{\r_K^\infty \r_\Xi^\infty} \over {\r_\p^\infty \r_\O^\infty}}
    \mbox{\hspace{1cm} and \hspace{1cm}}
    R_1 = {{\r_\p^\infty \r_\Xi^\infty}\over {\r_K^\infty \r_\O^\infty}} \;, 
\ee 
and $\r_h^\infty$ is the equilibrium number density of the hadron $h$. 
The $R_i$ quantities are the same as those in \cite{kmr}. They are
necessary because the thermal averaged cross-sections given in \Eref{eq:rd} 
are for the backward processes. These averaged cross-sections have been 
calculated in \cite{kmr} and their temperature dependence had been plotted 
in their Figure 5.2. One is seen to be weakly dependent on temperature
while the dependence of the other is quite strong. The latter is obviously
due to threshold because the reaction $\p + \Xi \longrightarrow K + \O$ has 
to overcome a mass gap of $\Delta m = m_{_\O}+m_{{_K}}-m_{_\Xi}-m_{_\p}$. 
For the purpose of obtaining the destruction timescales from the 
\Eref{eq:rd}, these thermal averaged cross-sections can be parameterized 
as follows.  
\be \langle \s_{_{ K + \Xi \rightarrow \p + \O}} v_{\p\O} \rangle    
      = 0.22 \;\; \mbox{mb $\cdot$ c}
\ee
\be \langle \s_{_{\p + \Xi \rightarrow  K + \O}}v_{ K\O}  \rangle 
    = 1.7 \Big (\frac{170}{T}\Big )\; \exp \{-705/T\} \; \mbox{mb $\cdot$ c} 
\ee
From them one can obtain the timescales for the two processes.  
\be \t_{_{ \p + \O \rightarrow  K + \Xi}} 
    = 160 \Big(\frac{170}{T}\Big )^{3/2} \exp \{142.5/T\} \; \mbox{fm/c}
\ee
\be \t_{_{  K + \O \rightarrow \p + \Xi}} 
    =  36 \Big(\frac{170}{T}\Big )^2     \exp \{354/T \}   \; \mbox{fm/c}
\ee
Putting in various values for the temperatures one finds
that these timescales, tabulated in \Tref{t}, are extremely long in the 
context of heavy ion collisions. The chance of $\O$ and $\aO$ produced
from DCC surviving the hadronic medium until the final breakup and thus
leaving a trace of DCC for detection is therefore very high.

\section{Summary and Remarks}

In this article signs of an anomaly in the $\O$ and $\aO$ produced
at the SPS were pointed out. These came in the form of a deviation
from the trend set by the other hadrons as well as from the yield
of the triply strange hyperon. There was an attempt to explain the anomaly 
in the slope of the $m_T$ spectra using the Relativistic Molecular Dynamics 
model in \cite{hsx}. Similar deviation in the linear increase of the
slope with mass was shown for $\O$. They concluded that the cause of this
was due to the fact that $\O$ and $\aO$ had not the same flow as the 
remaining hadrons. However they also found that $\O$ and $\aO$ were 
formed very early in the time evolution of the collisions. If there
was a formation of deconfined matter in these collisions, then the
$\O$ and $\aO$ would practically be existing at the same time with the
deconfined quarks and gluons. This point is quite difficult to understand.

We have proposed that DCC formation be the cause of the anomaly. 
The rarity of the yield of $\O$ and $\aO$ means that any addition
production mechanisms would readily manifest themselves and cause
deviation from the other more abundant hadrons.  Because $\O$ 
and $\aO$ thus produced would automatically be at low $p_T$ or 
essentially at rest, they naturally have not the same flow velocity as 
the other hadrons. Thus if we accept the flow related part of the 
explanation of the slope in the $m_T$ spectra provided by \cite{hsx}, 
a much smaller value of the slope parameter than expected is not too 
surprising. From the SPS data we have checked that our proposal is 
consistent with the non-observation of DCC in the distribution of low 
energy pion ratios and that the anomaly can survive the lifetime of the 
nuclear collisions. Therefore DCC formation could very plausibly be
responsible for the observed anomaly in $\O$ and $\aO$.

\ack
S.W. thanks the organizers for support, and for a very useful 
and enjoyable conference.

\end{document}